\documentclass[reprint,twocolumn,prl,groupedaddress,superscriptaddress,preprintnumbers,amsmath,amsfonts,amssymb,floatfix,showpacs]
{revtex4}

\usepackage{graphicx}
\usepackage{dcolumn}
\usepackage{bm}
\usepackage{epsfig}
\usepackage[usenames,dvips]{color}
\usepackage[english]{babel}
\usepackage{multirow}

\begin{document}
\title{Ferromagnetic spin coupling 
of 2$p$-impurities in band 
insulators stabilized by intersite Coulomb interaction: Nitrogen-doped MgO}

\author{I.~Slipukhina}
\email[Corresponding author: ]{i.slipukhina@fz-juelich.de}
\author{Ph.~Mavropoulos}
\author{S.~Bl\"ugel}
\author{M.~Le\v{z}ai\'{c}}
\affiliation{{Peter Gr\"unberg Institut and Institute for Advanced Simulation, Forschungszentrum J\"ulich and JARA,  D-52425 J\"ulich, Germany}}

\date{\today}

\begin{abstract}
For a nitrogen dimer in insulating MgO, a ferromagnetic coupling between spin-polarized
$2p$-holes is revealed by calculations based on the density functional theory amended
by an on-site Coulomb interaction and corroborated by the Hubbard model. It is shown that the ferromagnetic coupling is facilitated by a T-shaped orbital arrangement of the $2p$-holes, which is in its turn controlled by an intersite Coulomb interaction due to the directionality of the $p$-orbitals. We thus conjecture that this interaction is an important ingredient of  ferromagnetism in band insulators with $2p$ dopants.
\end{abstract}


\maketitle

\label{sec:introduction}

Ferromagnetic (FM) insulators offer the potential for use as active barrier materials 
in novel spintronic tunneling devices. In the past years we have witnessed a new vista in engineering FM insulators
not by doping nonmagnetic insulators with transition-metal ions as is traditionally accomplished in 
diluted magnetic semiconductors \cite{Sato_2010,Picozzi_2006}, but by doping with $sp$-elements.   
The novel magnetic materials design
was encouraged by unexpected experimental observations of
room-temperature ferromagnetism in defective wide-gap
oxide semiconductors and insulators 
\cite{Venkatesan_2004,Hong_2006,Hong_2008,Elfimov_2002,Hu_2008}. 
This phenomenon (referred to as $sp$- or $d^0$-magnetism) was associated
with the partially filled $p$-states of the intrinsic defects like
cation/anion vacancies \cite{Elfimov_2002,Venkatesan_2004,Pemmaraju_2005} or first-row ($2p$)
dopants \cite{Elfimov_2007}.

Shortly after, numerous cation-deficient or N- and C-doped oxides
and sulfides were theoretically predicted to be FM at room temperature \cite{Kenmochi_2004, Pardo_2008,Pan_2007,Ivanovskii_2007,Long_2009}.
Among those, MgO is distinct as it is certainly the most important barrier material for magnetic tunnel junctions.
Recently, N-doped MgO films were experimentally shown to exhibit ferromagnetic properties upon thermal annealing \cite{Yang_2010}. 

It is commonly concluded that the FM interaction between the defects in $d^0$-magnets
is due to partially occupied spin-polarized defect states, that 
are sufficiently extended to provide an exchange interaction via the
double-exchange mechanism. Most likely, however, at low concentrations the impurity
$2p$ electrons experience a strong on-site Coulomb repulsion $U$ because
of their spatial localization \cite{Peng09}, leading to insulating behavior 
that changes the mechanism of exchange interaction and could weaken or change the sign of magnetic coupling. 

Correlation effects in defect-free oxides with partially filled oxygen $p$-shells were studied in Refs.~\cite{Mahadevan_2010,Ederer_2009,Winterlik_2009}.
The importance of electron correlations in the impurity $p$-states was
recently investigated using density functional theory (DFT) approaches
amended by an on-site Coulomb repulsion  \cite{Pardo_2008,Droghetti_2008} for the
example of N-doped MgO. Jahn-Teller-like (JT) distortions, captured by
'$+U$' \cite{Liechtenstein_1995} or self-interaction correction \cite{Pemmaraju_2007} 
schemes, were found to evoke an energy splitting between the occupied
and unoccupied nitrogen $2p$-states, increasing the localization of a
spin-polarized hole at one of the $2p$-orbitals by pushing it deeper
into the band gap and thus breaking the initial symmetry of the electronic state. 

In this Letter we present new insights into the role
of electron correlations in N-doped MgO by carrying out
DFT calculations where strong correlations are accounted for by the
GGA (generalized-gradient approximation \cite{Perdew_1996})$+U$ approach, 
that are conceptualized on the basis of a minimal 
Hubbard model. We demonstrate that the symmetry breaking
and the subsequent metal-insulator transition
occurs even
without JT distortions, and is thus electronically driven. More
importantly we find for the insulating state of
N-dimers, which are likely to be formed in the annealing process, a FM
spin coupling and a T-shaped $2p$-orbital arrangement (OA).  
Employing a Hubbard model we show that a weak
intersite Coulomb repulsion,
that is well accounted for by the DFT,
combined with the on-site Coulomb
interaction is responsible for this result.  While in {\it periodic}
strongly correlated $3d$ systems the favored spin- and orbital arrangement
is usually well explained by the Kugel-Khomskii model \cite{Kugel_1973}
which does not resort to the intersite repulsion, similar arguments applied
to N-N dimers in MgO yield an antiferromagnetic (AFM) coupling, in contradiction with
our first-principles results.
The nearest-neighbor FM interaction is not sufficient to explain the long-range FM order, but it is a necessary condition. Our aim is to shed some light on the physics driving this interaction.

We  utilized the full-potential
linearized augmented plane wave 
code {\tt FLEUR} \cite{FLEUR} in our GGA$+U$ calculations. Relaxations were performed with an on-site Coulomb energy  $U$=4.6~eV and Hund's
exchange $J$=1.2~eV applied
on the $2p$-states of O and N. The total energies were compared \cite{Supplementary} for a set of $U$-values
that ranges between  3 and 6~eV, reported for O $2p$-states in transition-metal
oxides from photoemission and Auger experiments
\cite{Chainani_1992}, and for two values of $J$, 0.6 and 1.2~eV (from Hund's exchange for atomic N and O \cite{Supplementary}, we expect the physically relevant $J$-values in a solid to fall within this range). Supercells of 64 atoms of host MgO, with
a Brillouin-zone $2\times2\times2$ $\textbf{k}$-mesh 
are used to study a single N impurity and N-N dimers at O sites. N-O-N ``dimers''
are treated in a 144-atomic supercell and sampled at the $\Gamma$-point. 
Different OAs are introduced by 
initiating a specific occupation of the $p$-orbitals of N
in the GGA+$U$ density-matrix \cite{Schick_1999}. 
Calculations are performed for all possible types of OA  
at FM and AFM spin alignment 
(12 different spin-orbital configurations in total) for N-N and N-O-N dimers.

\label{sec:Single_impurity}

\begin{figure}[htbp]
	\begin{center}
		\includegraphics[width=1.0\hsize]{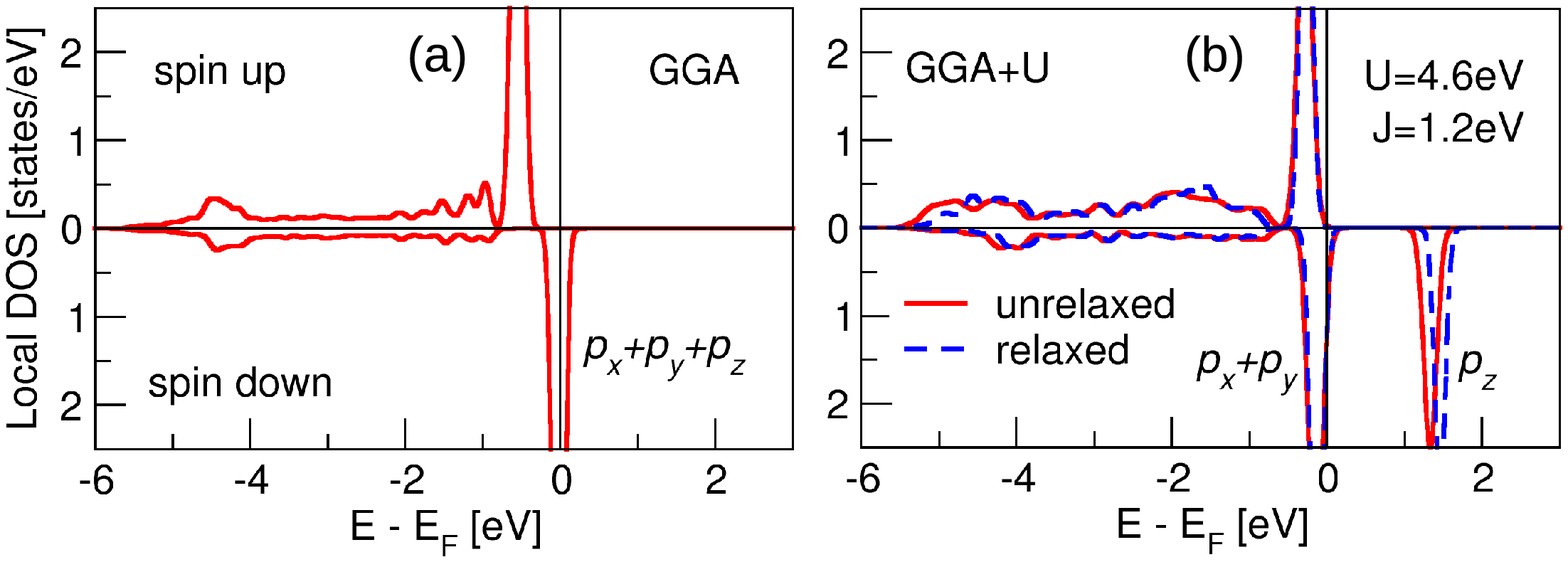}
		\includegraphics[width=1.0\hsize]{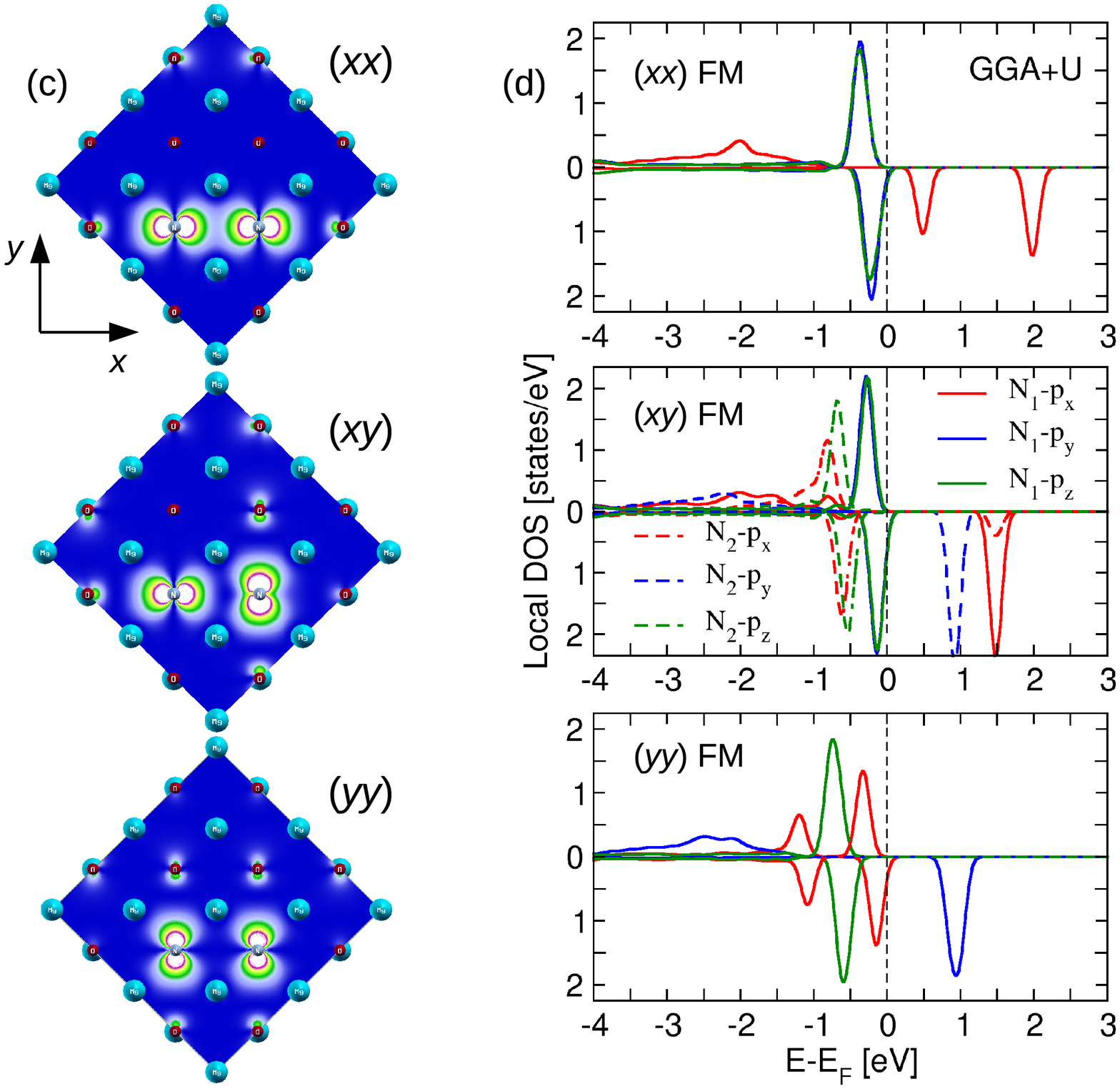}
 		\caption{(color online) Spin- and orbital-resolved DOS for a single N in MgO calculated in GGA (a) and GGA+$U$ (b). (c) Minority $p$-holes spin-density distribution in $xy$-plane around the N-N dimer, with ($xx$), T-shaped ($xy$),
 and ($yy$) OA. Blue, red, and grey spheres denote Mg, O, and N, respectively. (d) Spin- and orbital-resolved DOS of N-atom 1 (2) (solid (dashed) 
 lines) of the N-N dimer for the depicted OAs, with FM spin alignment; red, blue, green line: $p_x$-,  $p_y$-,  $p_z$-orbitals.\label{fig:Fig_1}}
	\end{center}
\end{figure}

We first consider the case of a single N impurity substituting O
in MgO to check whether the symmetry breaking and the metal-insulator transition upon applying   
a finite $U$ is driven electronically or by local lattice distortions. 
Figure~\ref{fig:Fig_1}(a) shows the N density of
states (DOS) of the structurally unrelaxed supercell. Within
the GGA, N introduces two triply degenerate states, an occupied 
spin-up $p$-state
in the valence band and a partially unoccupied 
spin-down state at the Fermi level, $E_{\rm F}$. A single spin-polarized
hole is evenly distributed among the three N $p$-orbitals.  The
situation changes drastically upon applying $U$: 
the minority $2p$-state splits into a doubly degenerate occupied and a
non-degenerate unoccupied level (Fig.\ref{fig:Fig_1}(b)). This symmetry breaking
occurs even without lattice distortions, which clearly demonstrates that it
is an electronically driven effect. Similar behavior was noticed for
the cation vacancy in ZnO \cite{Chan_2009}. This is
different from the previous reports on MgO:N \cite{Droghetti_2008}, where  the main effect of both
$+U$ or self-interaction corrections resulted in symmetry breaking via a JT
distortion.

\label{sec:Magnetic_and_orbital_ordering}

Next, we examine the relative spin and $p$-orbital arrangement for two
substitutional N impurities in MgO. Recent LSDA studies \cite{Wu_2010} have shown
that N-N impurity pairing in a structurally relaxed MgO supercell leads to a non-magnetic
insulating ground state with a fully occupied bonding $pp\sigma$ (as well as
$pp\pi$) and completely empty antibonding $pp\sigma^*$ states. However, we show that this is no longer the case in the presence of 
strong correlations. We consider the
following relative
positions of the two N atoms: (i) at the nearest neighbor (n.n.) sites, at a distance 
$d_\textrm{N-N}=2.97$~\AA~-- a N-N dimer; 
(ii) connected via an O atom ($d_\textrm{N-N}=5.83$~\AA~)
 -- a N-O-N dimer, and model, correspondingly, the 1st and 2nd n.n.\ 
interaction between N atoms in the oxygen sublattice of MgO. We omit
the N-Mg-N configuration from the analysis because we find the magnetic 
interactions to be negligibly small \cite{Mavropoulos_2009}. 

We define a coordinate system with the $p_x$ and
$p_y$ orbitals pointing toward neighboring N (or O) atoms in the
$xy$ plane (Fig.\ref{fig:Fig_1}(c)), while the $p_z$ orbital points out of plane, toward Mg. 
In the considered cases of N-N and N-O-N dimers, both N atoms are
situated along the $x$ axis, thus the largest hopping is of $pp\sigma$ type between neighboring $p_x$ orbitals, 
while $p_y$ and $p_z$ orbitals are essentially non-bonding. 
The sign and strength of the magnetic coupling between the two
spin-polarized holes is found by the total energy
difference $\Delta{E_{\rm AFM-FM}}$ between the AFM and FM state, calculated within GGA+$U$ for each $p$-hole OA. 
To be able to disentangle the electronic mechanism of OA from the JT
one, for the N-N dimer we performed calculations with and without
structural relaxations.

\begin{table}
 \caption{The GGA+$U$ total energies $\Delta{E}$ relative to the lowest-energy state and exchange interaction $\Delta{E_{\rm AFM-FM}}$ both in meV/supercell, calculated for different OA for N-N and N-O-N dimers in MgO. Positive (negative)  $\Delta{E_{\rm AFM-FM}}$ corresponds to FM (AFM) coupling.}\label{tab:table_1}
\begin{ruledtabular}
   \begin{tabular}{c|r@{ }lc|r@{ }lc|r@{ }lc}   
     & \multicolumn{3}{c|}{N-N}   & \multicolumn{3}{c|}{N-N, relaxed}  & \multicolumn{3}{c}{N-O-N}\\
    \hline
 OA 
 & \multicolumn{2}{c}{$\Delta{E}$} & $\Delta{E_{\rm AFM-FM}}$ & \multicolumn{2}{c}{$\Delta{E}$} &  $\Delta{E_{\rm AFM-FM}}$ & \multicolumn{2}{c}{$\Delta{E}$} &  $\Delta{E_{\rm AFM-FM}}$ \\ 
    \hline
    ($xx$)& \phantom{.}156&  & $-$129  &            99  & & $-$189 &  \phantom{1 }35 &  & $-$8 \\
    ($xy$)&            4  &  &  26     &            0   & &  34    &              7  &  & 2 \\ 
    ($xz$)&            0  &  &  24     &            13  & &  31    &              0  &  & 2 \\ 
    ($yz$)&            185&  &  1      & \phantom{.}258 & &   1    &              19 &  & 0 \\
    ($yy$)&            192&  &  2      &            280 & &   1    &              28 &  & 0 \\
    ($zz$)&            198&  &  5      &            294 & & 3.6    &              12 &  & 0 \\
    \end{tabular}
\end{ruledtabular}
\end{table}

The results are summarized in Table~\ref{tab:table_1}. For each OA the spin configuration (FM or AFM) with the lowest energy is shown. 
Clearly, there are substantial energy differences between different types of OA. 
Comparing the N-N and  N-O-N dimers, the rapid decrease of $\Delta{E_{\rm AFM-FM}}$ with the N-N distance  confirms previous studies \cite{Mavropoulos_2009}.
Importantly, 
for both dimers the FM T-shaped OA (($xy$) or ($xz$)) is of lowest energy, moreover for any choice of $U$ and $J$ parameters (see Supplemental Material \cite{Supplementary}).
 
It was suggested \cite{Droghetti_2008,Pardo_2008}
that the introduction of $U$ would reduce the magnetic coupling in ${d\rm^{0}}$-magnets.
However, a precise physical value of the parameter $U$  is not known.
To check the persistence of the magnetic coupling,
we repeated the calculations for the N-N dimer 
varying $U$ in the range between 0-6~eV (neglecting atomic relaxations), 
with $J/U$=0.26. At $U=J=0$, the system is half-metallic and ferromagnetic with
$\Delta{E_{\rm AFM-FM}}=265$~meV. However, at $U>2$~eV, it becomes 
insulating and the interaction through the double-exchange mechanism is no longer possible.
Thus, the magnetic coupling is much weaker
and we expect AFM coupling due to kinetic exchange interaction.
Instead, orbital arrangement sets in and despite the insulating state, the interaction between the
spin-1/2 holes remains FM with total magnetic spin moment of $2\mu_{\rm B}$.
In the following we 
provide a qualitative understanding of this unexpected
finding. 

\begin{figure}[htbp]
	\begin{center}
\includegraphics[width=0.95\hsize, angle=0]{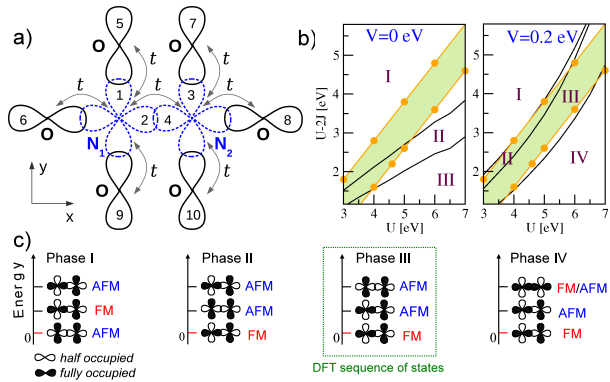}
\caption{(color online) 
(a) A schematic view of the cluster of $p$-orbitals, used in the model Hamiltonian (\ref{equ_1}). The dashed blue line depicts $p_x$- and $p_y$-orbitals of the n.n. nitrogen atoms (N$_1$ and N$_2$), while the solid black line depicts the corresponding orbitals of the O atoms. The possible hopping processes $t$ between two orbitals on neighboring sites are also shown. (b) The ($U-2J$) vs.\ $U$ phase diagram of the model for $V=0$ and $V=0.2$~eV. Orange circles denote the $U$ and $J$ values, considered in our {\it ab initio} calculations \cite{Supplementary}, outlining the  physically relevant area (in green). (c) The phases I, II, III and IV, corresponding to different sequences of several orbitally-arranged states in the order of energetic stability. }\label{fig:Fig_2}
	\end{center}
\end{figure}

In the first step we show that kinetic exchange alone favors an AFM ground state. We consider 
a minimal multiorbital Hubbard model
that describes the system of strongly correlated open-shell
$2p$-electrons in terms of the kinetic energy, $t$, of 
spin-conserving hoppings between orbitals of largest overlap 
(n.n.\ $pp\sigma$-orbitals, see Fig.\ref{fig:Fig_2}(a)), the on-site
Coulomb repulsion energy $U$ and the interorbital exchange (Hund's rule) 
coupling $J$ (see also Eq.~(\ref{equ_1})).

For a given OA, the favored magnetic interaction is determined by hopping of the electrons within the N-N-dimer.
A substantial magnetic coupling occurs only for 
those OAs that allow hopping between the two $p_x$-orbitals of the 
N-dimer, where at least one of these orbitals is half-filled: 
the ($xx$) and the T-shaped ($xy$) and ($xz$)
OAs (see Table \ref{tab:table_1}).
The relative coupling strength can be qualitatively understood by considering that each
spin-conserving hopping $t$ into half-filled orbitals lowers the
energy by $t^2/(U-J)$ (in 2nd order perturbation theory) if the
intermediate virtual atomic state satisfies Hund's rule for one of the
atoms, and $t^2/U$ otherwise. In the ($xx$) OA, this gives
$E_{\rm AFM}=-2t^2/U$, $E_{\rm FM}=0$, \textit{i.e.} an AFM ground state 
(the factor 2 arises because there is
one half-filled 
orbital per atom participating in the hopping). In the
T-shaped OA there is only one half-filled 
orbital participating; then we have $E_{\rm AFM}=-t^2/U$, $E_{\rm FM}=-t^2/(U-J)$, \textit{i.e.}, a FM
ground state. 

Among all possible types of OA, the T-shaped and AFM ($xx$) states 
are the ones that have the lowest energy, because
these are the only ones that allow hopping between the N atoms.
The hopping between O and N atoms provides a smaller total energy gain, because of
(i) the separation between the on-site energy
levels of O and N, that differ 
by 0.5-1 eV, and (ii) the delocalization of the O states that form an
itinerant band.  From the previous discussion follows
that among the two OAs, the AFM ($xx$) one should be lower in 
energy by $2t^2/U-t^2/(U-J)$ as long as $J<U/2$.  
Thus, the arguments of the kinetic exchange interaction favor the AFM ($xx$) state, contradicting our DFT finding of a FM T-shaped ground state. 

\label{sec:Model_Hamiltonian}
Obviously, a qualitative description of the relative stability of spin- and orbitally-arranged 
states in N-doped MgO requests an extension of the minimal Hubbard model
beyond the conventional electron hopping $t$ and \textit{on-site} Coulomb repulsion $U$.
The N-N dimer breaks the cubic symmetry of the MgO lattice. 
Considering the directional nature of the $2p$-orbitals, 
this leads to an orbital-dependent
\textit{intersite} Coulomb repulsion, that is well accounted for  within the GGA/LDA. 
In terms of the Hubbard model it is expressed
by an additional effective intersite Coulomb repulsion energy $V$ between n.n.\ $pp\sigma$-orbitals of N-N and N-O pairs. To capture this effect by such an  extended
Hubbard model we study an impurity 
cluster of 2 N and 6 n.n.\ O atoms (Fig.\ref{fig:Fig_2}(a)), with $1$ ($2$)  
$p$-orbital on each O (N) site that form n.n.\ $pp\sigma$-orbitals.
Thus, we 
restrict ourselves to 10 $p$-orbitals that can be occupied by 18 electrons 
and 2 holes of spins $\uparrow$ and $\downarrow$. 
The  Hamiltonian for such a system can be written as:
\begin{eqnarray}
H&=&\sum_{m,s}{\epsilon_{m}}{n_{ms}}-t\!\sum_{<m,m'>}\sum_{s}[c_{ms}^{\dag}c_{m's}+h.c.] \nonumber \\
&+&U\sum_{m}{n_{m\uparrow}n_{m\downarrow}}+{V}\!\!\!\sum_{<m,m'>}\!\!\!\!{n_{m}n_{m'}} \label{equ_1} \\
&+&(U'-J)\!\!\!\!\!\!\sum_{\ll m,m'\gg ,s}\!\!\!\!\!n_{ms}n_{m's}+ U'\!\!\!\!\!\sum_{\ll m,m'\gg ,s}{\!\!\!\!\!n_{ms}n_{m'-s}} \nonumber
\end{eqnarray}
Here $<\!\!m,m'\!\!>$ indicates the pair of n.n.\ orbitals of $pp\sigma$-overlap, 
$\epsilon_{m}$ denotes the energy level, $c_{ms}^{\dag}$ 
($c_{ms}$) creates (annihilates) a particle of spin 
$s=\uparrow,\downarrow$ of orbital $m$, while
$n_{ms}$, $n_{m}=n_{m\uparrow}+n_{m\downarrow}$ are the spin-number and
number operators, respectively.
Since we have two orbitals at the N atoms, $U'$ accounts for the
\textit{inter}-orbital on-site Coulomb interaction;
$\ll\!\! m,m'\!\!\gg$ stands for the two on-site orbital pairs (1,2) and (3,4).
We neglect 
on-site spin-flip and pair-hopping terms as they are non-zero only for the states with an empty 
and a fully occupied orbital at the same site, which 
turned out to be of much higher energy. 
The matrix elements of (\ref{equ_1}) are evaluated in the
18-electron-state basis (190 basis functions in total). The
eigenvalues of the model are obtained numerically by exactly diagonalizing
(\ref{equ_1}) for a given set of parameters. We impose the relations 
$U>U'>J>0$, $U'=U-2J$ (for a derivation see the Supplemental Material \cite{Supplementary}) and we take $t=0.75$~eV (extracted from the
splitting of the DOS on Fig.\ref{fig:Fig_1}(d)). 
As  the $p$-levels of
N atoms are higher in energy than those of O, we take
$\epsilon_{m}=0.5$~eV for N and zero for O. The $p_z$-orbitals which
are omitted from the model are considered to be always filled
and have only an effect of a constant shift on the on-site energies
$\epsilon_{m}$.

Figure~\ref{fig:Fig_2}(b) shows a ($U-2J$) vs.\ $U$ phase diagram, resulting from Hamiltonian (\ref{equ_1}) for $V=0$ and $V=0.2$~eV. The black lines denote the boundaries between the phases I, II, III and IV, corresponding to different sequences  of energetic stability of OAs (see Fig.~\ref{fig:Fig_2}(c)).  
We find that the FM ($xy$)
state is the lowest-energy one for the phases II, III and IV. However, for almost all $U$ and $J$ values, the AFM ($xx$) state (phase I) is the ground state if $V=0$, contradicting our {\it ab initio} results \cite{Supplementary}. Moreover, at $V=0$  the sequence of DFT states from Table~\ref{tab:table_1} (phase III) is not reproduced. In contrast, at finite $V$ the whole sequence of DFT states is reproduced for a wide range of physically relevant values of $U$ and $J$. Hence, the intersite Coulomb interaction $V$, which defines the repulsion of electrons on n.n. $p$-orbitals directed towards each other, is crucial for stabilizing the FM state in doped insulators with partially occupied $p$-orbitals like MgO:N.

\label{sec:conclusions}

In summary, we have shown
that the physics of spin-polarized holes of N-doped 
MgO is largely determined by $p$-electron correlation effects. 
For a single impurity, the splitting of the N $2p$-states at finite $U$ 
is mainly an electronic effect, enhanced by the lattice distortions. 
Moreover, the $p$-electron correlations lead to a spin and orbital arrangement of the $2p$-hole states of the N-N dimer, resulting in a FM coupling between T-shaped orbitally-arranged spin-polarized $p$-holes. The
FM state is realized
at reasonable values of $U$ and $J$, if an {\it intersite} Coulomb repulsion $V$ is accounted for. Considering the small value of $V\approx 0.2$~eV, necessary to drive the phase transition from an AFM to a FM ground state, and the fact that the directional nature of the $p$-orbitals 
controls the strength of
the intersite Coulomb repulsion, we conjecture that our finding affects all oxides where magnetism is due to $p$-electrons or defects in the $p$-electron systems.

\label{sec:acknowledgement}
We appreciate valuable discussions with K.~Rushchanskii and A.~Liebsch, 
and the support of the J\"ulich Supercomputing Centre. I.S.\ and M.L.\ gratefully acknowledge the support of the Young Investigators Group Program of the Helmholtz Association, Contract VH-NG-409. 

\label{sec:bibliography}


\begin{thebibliography}{10}

\bibitem{Sato_2010}
K.~{Sato}, L.~{Bergqvist}, J.~{Kudrnovsk\'y}, P.H.~{Dederichs}, O.~{Eriksson}, I.~{Turek}, B.~{Sanyal}, G.~{Bouzerar}, H.~{Katayama-Yoshida}, V.A.~{Dinh}, T.~{Fukushima}, H.~{Kizaki}, R.~{Zeller}, 
\newblock{Rev. Mod. Phys.} \textbf{82},1633 (2010).

\bibitem{Picozzi_2006}
S.~{Picozzi}, M.~{Le\v{z}ai\'{c}}, S.~{Bl\"{u}gel},  
\newblock{phys. stat. sol. (a)} \textbf{203}, 2738 (2006).

\bibitem{Venkatesan_2004}
M.~{Venkatesan}, C.B.~{Fitzgerald}, J.M.D.~{Coey}, 
\newblock {Nature (London)} \textbf{430}, 630 (2004).

\bibitem{Hong_2006}
N.H.~{Hong}, J.~{Sakai}, N.~{Poirot}, V.~{Briz\'e}, 
\newblock {Phys. Rev. B} \textbf{73}, 132404 (2006).

\bibitem{Hong_2008}
N.H.~{Hong}, N.~{Poirot}, J.~{Sakai}, 
\newblock {Phys. Rev. B} \textbf{77}, 033205 (2008).

\bibitem{Elfimov_2002}
I.S.~{Elfimov}, S.~{Yunoki}, G.A.~{Sawatzky},
\newblock {Phys. Rev. Lett.} \textbf{89}, 216403 (2002).

\bibitem{Hu_2008}
J.~{Hu}, Z.~{Zhang}, M.~{Zhao}, H.~{Qin}, L.~{Sun}, X.~{Kong}, M.~{Jiang},
\newblock {Appl. Phys. Lett.} \textbf{93}, 192503 (2008).

\bibitem{Pemmaraju_2005}
C.~{DasPemmaraju} and S.~{Sanvito},
\newblock {Phys. Rev. Lett.} \textbf{94}, 217205 (2005).

\bibitem{Elfimov_2007}
I.S.~{Elfimov}, A.~{Rusydi}, S.I.~{Csiszar}, Z.~{Hu}, H.H.~{Hsieh}, H.-J.{Lin}, C.T.~{Chen}, R.~{Liang}, G.A.~{Sawatzky},
\newblock {Phys. Rev. Lett.} \textbf{98}, 137202 (2007).

\bibitem{Kenmochi_2004}
K.~{Kenmochi}, M.~{Seike}, K.~{Sato}, A.~{Yanase}, H.~{Katayama-Yoshida},
\newblock {Jpn. J. Appl. Phys.} \textbf{43}, L936 (2004).

\bibitem{Pardo_2008}
V.~{Pardo} and W.E.~{Pickett}, 
\newblock {Phys. Rev. B.} \textbf{78}, 134427 (2008).

\bibitem{Pan_2007}
H.~{Pan}, J.B.~{Yi}, L.~{Shen}, R.Q.~{Wu}, J.H.~{Yang}, J.Y.~{Lin}, Y.P.~{Feng}, J.~{Ding}, L.H.~{Van}, J.H.~{Yin},
\newblock {Phys. Rev. Lett.} \textbf{99}, 127201 (2007).

\bibitem{Ivanovskii_2007}
A.L.~{Ivanovskii},
\newblock {Physics Uspekhi}, \textbf{50}, 1031 (2007).

\bibitem{Long_2009}
R.~{Long} and N.J.~{English},
\newblock {Phys. Rev. B} \textbf{80}, 115212 (2009).

\bibitem{Yang_2010}
C.-H.~{Yang},  
\newblock {PhD thesis, Stanford University} (2010).

\bibitem{Peng09}
H.~{Peng}, H.J.~{Xiang}, S.-H.~{Wei}, S.-S.~{Li}, J.-B.~{Xia}, J.~{Li},
\newblock{Phys. Rev. Lett.} \textbf{102}, 017201 (2009).

\bibitem{Mahadevan_2010}
A.K.~{Nandy}, P.~{Mahadevan}, P.~{Sen}, and D.D.~{Sarma}, 
\newblock {Phys. Rev. Lett.} \textbf{105}, 056403 (2010).

\bibitem{Ederer_2009}
R.~{Kov\'a\v{c}ik}, and C.~{Ederer}, 
\newblock {Phys. Rev. B} \textbf{80}, 140411 (2009).

\bibitem{Winterlik_2009}
J.~{Winterlik}, G.H.~{Fecher}, C.A.~{Jenkins}, C.~{Felser}, C.~{M\"uhle}, K.~{Doll}, M.~{Jansen}, L.M.~{Sandratskii}, and J.~{K\"ubler},
\newblock {Phys. Rev. Lett.} \textbf{102}, 016401 (2009).

\bibitem{Droghetti_2008}
A.~{Droghetti}, C.D.~{Pemmaraju}, S.~{Sanvito}, 
\newblock {Phys. Rev. B.} \textbf{78}, 140404(R) (2008).

\bibitem{Liechtenstein_1995}
A.I.~{Liechtenstein}, V.I.~{Anisimov}, J.~{Zaanen}, 
\newblock {Phys. Rev. B} \textbf{52}, R5467 (1995).

\bibitem{Pemmaraju_2007}
C.D.~{Pemmaraju}, T.~{Archer}, D.~{Sanchez-Portal}, S.~{Sanvito},
\newblock {Phys. Rev. B}, {\bf 75}, 045101 (2007).

\bibitem{Perdew_1996}
J.~P. {Perdew}, K.~{Burke}, M.~{Ernzerhof},
\newblock {Phys. Rev. Lett.} \textbf{77}, 3865 (1996).

\bibitem{Kugel_1973}
 K.I.~{Kugel}, D.I.~{Khomskii},,
\newblock {Zh. Eksp. Teor. Fiz.} \textbf{64}, 1429 (1973)
\newblock {[Sov. Phys. JETP} \textbf{37}, 725 (1973)].

\bibitem{FLEUR}
http://www.flapw.de/

\bibitem{Supplementary}
See EPAPS Document No. ?????? for
Supplemental Material. For more information on EPAPS,
see http://www.aip.org/pubservs/epaps.html.

\bibitem{Chainani_1992}
 A.~{Chainani}, M.~{Mathew}, D.D.~{Sarma},
\newblock {Phys. Rev. B} \textbf{46}, 9976 (1992).

\bibitem{Schick_1999}
 A.B.~{Shick}, A.I.~{Liechtenstein}, W.E.~{Pickett},
\newblock {Phys. Rev. B} \textbf{60}, 10763 (1999).

\bibitem{Chan_2009}
 J.A.~{Chan}, S.~{Lany}, A.~{Zunger},
\newblock {Phys. Rev. Lett.} \textbf{103}, 016404 (2009).

\bibitem{Wu_2010}
 H.~{Wu}, A.~{Stroppa}, S.~{Sakong}, S.~{Picozzi}, M.~{Scheffler}, P.~{Kratzer},
\newblock {Phys. Rev. Lett.} \textbf{105}, 267203 (2010).

\bibitem{Mavropoulos_2009}
 Ph.~{Mavropoulos}, M.~{Le\v{z}ai\'{c}}, S.~{Bl\"{u}gel},
\newblock {Phys. Rev. B} \textbf{80}, 184403 (2009).


\end{thebibliography}
\end{document}